\documentclass[aps,preprint,showkeys,groupedaddress]{revtex4}

\usepackage{graphicx}
\usepackage{amsmath}
\usepackage{amssymb}
\begin{document}

\title{Molecular packing and chemical association in liquid water simulated using {\em ab initio} hybrid Monte Carlo and different exchange-correlation functionals}

\author{Val\'ery Weber}
\thanks{Email: vweber@unizh.ch}
\affiliation{Physical Chemistry Institute, University of Zurich, 8057 Zurich, Switzerland.}%
\author{Safir Merchant}
\author{Purushottam D. Dixit}
\author{D. Asthagiri}
\thanks{Email: dilipa@jhu.edu; Fax: +1-410-516-5510}
\affiliation{Department of Chemical and Biomolecular Engineering, Johns Hopkins University, Baltimore, Maryland 21218, USA.}%

\date{\today}

\begin{abstract}
In the free energy of hydration of a solute, the chemical contribution is given by the free energy required to expel water molecules from the coordination sphere  and the packing contribution  is given by the free energy required to create the solute-free coordination sphere (the observation volume) in bulk water.  With the SPC/E water model as a reference, we examine the chemical and packing contributions in the free energy of water simulated using different electron density functionals. The density is fixed at a value corresponding to that for SPC/E water at a pressure of 1~bar.  The chemical contribution shows that water simulated at 300~K with BLYP is somewhat more tightly bound than water simulated at 300~K with the revPBE functional or at 350~K with the BLYP and BLYP-D functionals.  The packing contribution for various radii of the observation volume is studied. In the size range where the distribution of water molecules in the observation volume is expected to be Gaussian,  the packing contribution is expected to scale with the volume of the observation sphere. Water simulated at 300~K with the revPBE and at 350~K with BLYP-D or BLYP conforms to this expectation, but the results suggest an earlier onset of system size effects in the BLYP 350~K and revPBE 300~K systems than that observed for either BLYP-D~350~K or SPC/E.  The implication of this observation for constant pressure simulations is indicated. For water simulated at 300~K with BLYP, in the size range where Gaussian distribution of occupation is expected, we instead find non-Gaussian behavior, and the packing contribution scales with surface area of the observation volume, suggesting the presence of heterogeneities in the system. 
\end{abstract}
\maketitle

\section{Introduction}

The structure of nonassociated liquids such as liquid nitrogen or liquid argon can be understood in terms of 
packing of molecules \cite{chandler:wca83}: the structure is primarily determined by hard-core repulsive (excluded-volume) interactions and not by  specific, directional intermolecular interactions. 
Further, the thermodynamics of such fluids admits  a mean-field (van der Waals type) approximation \cite{chandler:wca83} (see also Fig.~1 in Ref.\ \cite{shah:jcp07}). 
In contrast,  for an associated liquid like water, the attractive interactions are strong and specific, revealing themselves, for example, in the approximately tetrahedral ordering of water molecules around a central water molecule \cite{stillinger:sc80}. In such a case, the observed structure and thermodynamics reflects both packing interactions and local, chemically specific interactions between molecules. To understand the structure and thermodynamics of liquid water, it is thus imperative to understand the balance between packing effects and local, attractive interactions.

Recent developments in molecular statistical thermodynamics \cite{lrp:apc02,lrp:book,asthagiri:pre03,paliwal:jcp06,asthagiri:jacs2007,shah:jcp07,asthagiri:jcp2008}  
allow a detailed examination of the competing roles of packing and local, chemically involved interactions in 
the physics of hydration.  These efforts are founded on regularizing\cite{asthagiri:jcp2008} the statistical problem of calculating the excess chemical potential of the solute using the potential distribution theorem\cite{lrp:book,widom:jpc82}. 
By introducing a spatial scale --- here the coordination radius of interest --- the interaction of the solute with the solvent 
is separated into a local, chemically interesting piece and a long-range piece. The coordination radius can be adjusted such that  the distribution of 
energies from the long-range interaction piece admits a simplified description \cite{asthagiri:jacs2007,shah:jcp07,asthagiri:jcp2008}. This then helps focus the attention on the local problem
solely. 

The local contributions to hydration are accounted for by the work of expelling the water molecules from within the coordination sphere in two limits, one in the presence of solute-solvent interactions and the other with those interactions turned off.  The 
former gives the chemical contributions to hydration whereas the latter accounts for packing
contributions. (We will refer to the coordination sphere without the solute as the observation volume.) From a simulation record, the chemical contribution can be obtained by noting the probability, $x_0$, of observing no water molecules in the coordination sphere. Likewise, the packing contribution is obtained from the probability, $p_0$, of observing an empty observation volume --- a cavity --- in bulk water. The $x_0$ and $p_0$ values are, respectively, the $n=0$ members of the set $\{x_n\}$ and $\{p_n\}$ of occupancy number ($n$) distributions in the coordination sphere and the observation volume.
  
For coordination radii that are chemically meaningful for the solute of interest, the distribution 
of coordination states $\{x_n\}$ below the most probable coordination state indicate the relative contribution of those
states to the local, chemically involved contributions to hydration \cite{merchant:jcp09}. Likewise the distribution
$\{p_n\}$ provides important insights into the hydrophobic aspects of hydration \cite{Hummer:1996p326,garde:prl96,lrp:jpcb98}. 
For observation volumes with radii in the range 2.0-3.5~{\AA}, $\{p_n\}$  is found to be nearly Gaussian \cite{paliwal:jcp06,Hummer:1996p326,garde:prl96,lrp:jpcb98}.  In this case, the variance of $\{p_n\}$, and hence also the excess chemical potential of the cavity,  is expected to scale linearly with the volume of the cavity \cite{chandler:nature05}. (This scaling
relation proves insightful in the analysis below.) 

Earlier, for a classical, empirical model of liquid water, based on an analysis of the $\{x_n\}$ and $\{p_n\}$ distributions 
it was found that the packing and chemical contributions balance at a coordination radius of about 3.3~{\AA}  \cite{shah:jcp07,paliwal:jcp06}. At that point, the net hydration free energy is entirely determined by non-specific interactions between the water molecule and the bulk liquid outside the coordination sphere \cite{shah:jcp07}. A similar conclusion was reached for liquid water simulated with the revPBE functional within constant $NVE$ {\em ab initio} molecular dynamics \cite{asthagiri:pre03}. In that study, because of limited data, using the more robustly determined mean and variance 
of the number distributions, a maximum entropy approach was used to secure $x_0$ and $p_0$ \cite{asthagiri:pre03}. Further, 
relative to water simulated with the revPBE functional, it was found that the enhanced structure obtained using the PBE functional correlated with attractive interactions outweighing packing effects.

There were two main limitations in the earlier {\em ab initio\/} molecular dynamics study \cite{asthagiri:pre03}. First, the simulations with the revPBE and PBE functionals were at different average temperatures (314~K and 337~K, respectively), 
confounding any clear comparison between the functionals. Second, the total simulation time was small, being less than 15~ps for any one functional under study. In the present work, we address both those limitations. We implement a hybrid Monte Carlo (HMC) method \cite{duane:plb87,creutz:prd88,gupta:plb90,mehlig:prb92,smit}, an approach that guarantees that we sample from a canonical ensemble without in any way influencing the forces obtained using the functional under study. The simulations are also conducted for a longer time (about 170~ps).

In the following section, we briefly summarize the statistical thermodynamic theory and recapitulate main ideas of the HMC method. In Section~\ref{sc:methods}  we outline the methods used. For the {\em ab initio} simulations, we implement the HMC method as a script that interfaces with the publicly available {\sc CP2K} code  \cite{cp2knew}. In Section~\ref{sc:results} we present the results of our study.  For water simulated with BLYP, the packing contribution shows a scaling behavior that is not expected of a homogeneous liquid material, whereas it does so for water simulated with revPBE at 300~K and with BLYP and BLYP-D at 350 K.  With scaled particle theory as a reference \cite{ashbaugh:rmp}, these results suggest
that the density (pressure) of the liquid simulated with BLYP at 350~K and revPBE at 300~K is somewhat higher than normal. 
Overall, among the models studied here, water simulated with BLYP-D at 350~K appears to best describe liquid water at 300~K. 

\section{Theory}
\subsection{Statistical thermodynamic theory}
Consider a solute $\rm X$ in a bath of water molecules  and define a coordination sphere  of 
radius $\lambda$ around the solute. In the $n$-coordinate state of the solute, there are exactly $n$ water-oxygens
in the coordination sphere. The probability of the $n$-coordinate state, $x_n$, is the fraction of  observing such cases. 
In the absence of the solute-solvent interactions, the probability of $n$ water-oxygen atoms to be found within 
the observation volume is $p_n$. (Recall that the observation volume refers to the coordination sphere without the solute.) The probabilities $x_n$ and $p_n$ are related by \cite{merchant:jcp09,asthagiri:cpl10} 
\begin{eqnarray}
x_n = p_n e^{-\beta [ \mu_{\rm X}^{\rm ex}(n) - \mu_{\rm X}^{\rm ex}]} \, ,
\label{eq:xnpn}
\end{eqnarray}
where $\beta^{-1} = k_{\rm B} T$ and $T$ is the temperature and $k_{\rm B}$ is the Boltzmann constant. The excess chemical potential of the solute, $\mu_{\rm X}^{\rm ex}$, is the chemical potential in excess of the ideal gas result at the same density and temperature of the solute $\rm X$ in the solution, and $\mu_{\rm X}^{\rm ex}(n)$ is the excess chemical potential
with the added constraint that 
only $n$-water molecules are present in the coordination sphere.  Observe that in Eq.~\ref{eq:xnpn}, 
$p_n$ defines the intrinsic propensity of solvent molecules to populate the observation volume. Solute-solvent interactions codified by the quantity $\mu^{\rm ex}_{\rm X}(n) - \mu^{\rm ex}_{\rm X}$ modify this intrinsic or bare probability 
to give $x_n$.

Eq.~\ref{eq:xnpn} can be rigorously established \cite{merchant:jcp09,asthagiri:cpl10} --- here it suffices to note that
the normalization of probabilities $\{x_n\}$ and $\{p_n\}$ lead to well-established multi-state generalizations of the potential distribution theorem \cite{merchant:jcp09,asthagiri:cpl10,lrp:mulGjacs97,lrp:book}. The physical content of Eq.~\ref{eq:xnpn} is best seen for the $n=0$ case: 
\begin{equation}
\beta \mu^{\rm ex}_{\rm X} = \ln x_0 - \ln p_0 + \beta \mu^{\rm ex}_{\rm X} (n=0) .
\label{eq:xopo}
\end{equation}
The packing contribution, $\beta \mu_{\rm HS} = -\ln p_0$, is the free energy of forming an empty observation volume of a defined radius, a hard sphere (HS), in bulk water. The interaction free energy of the solute in the center of an empty 
coordination sphere is $\beta \mu^{\rm ex}_{\rm X} (n=0)$; the operation is one of placing the solute at the center of the 
cavity. Finally, the free energy change in allowing water to flood the empty coordination sphere is $\ln x_0$; thus this term measures the role of the specific, directional bonding between the solute and the solvent within the
coordination sphere. (The relation $x_0 = 1/ \sum K_{n} \rho_w^n$, where $K_n$ is the equilibrium constant for forming a solute plus $n$-water cluster within the coordination sphere and $\rho_w$ is the density of water \cite{lrp:apc02,lrp:book,paliwal:jcp06,merchant:jcp09,lrp:cpms}, provides an alternative perspective on the chemical contributions 
to $x_0$.)

\subsection{Hybrid Monte Carlo}\label{sc:hmc}

The hybrid Monte Carlo (HMC) method combines elements of Monte Carlo and molecular dynamics approaches and has been well-established in the literature \cite{duane:plb87,creutz:prd88,gupta:plb90,mehlig:prb92,smit}.  We briefly recapitulate the main ideas of this method here. 

We assume, as is the case here, that the $N$-particle system is described by a classical 
Hamiltonian $\mathcal{H}(q,p) = \mathcal{T}(p) + \mathcal{U}(q)$. $\mathcal{T}$ and $\mathcal{U}$ 
are, respectively, the usual kinetic and potential energy, and $q$ are the 
coordinates and $p$ the conjugate momenta.  $\mathcal{U}(q)$ is obtained using electron density
functional theory, and we are interested in the canonical distribution of configurations of this system. 

The canonical configurations of this system can be generated by conventional (Metropolis)
Monte Carlo: we make a one-particle move and accept or reject that move based on the 
standard Metropolis criterion. The Metropolis criterion also specifies the temperature
of the system.  To better explore configurational space it would be desirable to make 
multiple moves at a time, but such an approach is rather inefficient in conventional 
Monte Carlo. The HMC method removes this inefficiency by combining 
the effectiveness of molecular dynamics to generate $N$-particle (global) moves with the
effectiveness of Monte Carlo in rigorously generating a canonical distribution \cite{duane:plb87,creutz:prd88,gupta:plb90,mehlig:prb92,smit}. 

One sweep of the HMC consists of $N_{md}$ molecular dynamics time steps of propagation through phase space starting from the configuration $(q,p)$. The dynamics are performed at constant energy using a time reversible and area preserving discretization scheme  and a time-step of $\delta t$.
Initial momenta $p$ are assigned from a Gaussian distribution at the inverse temperature $\beta$. 
At the end of $N_{md}$ time-steps a candidate phase space configuration $(q',p')$ is generated.  
If  $\delta \mathcal{H} = \mathcal{H}(q',p') - \mathcal{H}(q,p)$ is the discretization error, then 
the new configuration is accepted with a probability 
\begin{eqnarray}
P_A  = {\rm min}\{1, e^{-\beta \delta \mathcal{H}}\} \; . \label{eq:pa}
\end{eqnarray}
The above procedure guarantees that detailed balance is satisfied \cite{duane:plb87,creutz:prd88,gupta:plb90,mehlig:prb92}. 

 There are two helpful properties of the hybrid Monte Carlo method that can serve as 
 consistency checks of the simulation.  First, the area preserving property implies that
 \cite{gupta:plb90,mehlig:prb92}
 \begin{eqnarray}
 \langle e^{-\beta \delta \mathcal{H}}\rangle = 1 , 
\label{eq:norm1} 
\end{eqnarray}
where $\langle \ldots\rangle$ denote a canonical average. Second, provided the third
and higher cumulants of Eq.~\ref{eq:norm1} vanish, as happens when $N$ is large and $\delta t$ is small
such that the variance of the distribution of $\delta\mathcal{H}$ is a constant, 
then \cite{gupta:plb90,mehlig:prb92}
\begin{eqnarray}
P_A = {\rm erfc}(\frac{1}{2}\sqrt{\beta\delta\mathcal{H}}) \; . \label{eq:erfc}
\end{eqnarray}

In the hybrid Monte Carlo procedure, the acceptance rate depends on $\delta t$ and $N_{md}$.
 Here, in the initial phase of the simulation, we fix $N_{md}$ and 
block average data every 10 sweeps to determine the acceptance ratio. If the acceptance ratio is greater (lesser) than the targeted ratio, $\delta t$ is increased (decreased) by 10\% to target the specified acceptance ratio. In the next phase, 
the time step $\delta t$ is held constant, ensuring that strict detailed balance is satisfied. 
Only results from the latter phase are reported here.

\section{Methodology}\label{sc:methods}

\subsection{Classical MC simulation}
The SPC/E water model \cite{spce} was taken as a reference for the
structure and coordination number distributions of principal interest in this work. This choice was motivated by several considerations: (1) we seek coordination number distributions, and hence using a model 
is inevitable; (2) the oxygen-oxygen pair-correlation function, $g_{\rm OO}(r)$, obtained using SPC/E is in reasonable agreement with the current best experimental results using the Advanced Light Source \cite{headgordon:cr02} experiment (see below); and (3) the SPC/E model is known to well-describe the liquid-vapor phase boundary of water \cite{guillot:waterjcp93,guillot:jml02} (see also Refs.\ \cite{vega:prl04} and \cite{vega:faraday09}).

A 32 water molecule system at a number density $\rho = 33.33$~{nm}$^{-3}$ (mass density
of 0.997~g/cm$^3$) corresponding to 1~bar pressure \cite{vega:prl04} was simulated using Metropolis Monte Carlo \cite{allen}. Long-range electrostatic interaction was described using Ewald summation and  Lennard-Jones interactions were truncated at half the box-length. After an initial equilibration of $3\times 10^5$ sweeps, where one sweep is one attempted move per particle, data collected over an additional $3\times 10^5$ sweeps was used for analysis. The structure and thermodynamics from the small system are in excellent agreement with results using larger systems (data not shown).

\subsection{Classical HMC simulations}
For our initial studies with the HMC method, we simulated the classical, flexible SPC/Fw \cite{voth:spcfw06} model of water.
(We consider HMC simulations with the flexible model, as this is of  most interest in {\em ab initio\/} simulations.) 
 The HMC method was implemented using a Python script
that computes $\delta \mathcal{H}$ between $N_{md}$ steps of dynamics and appropriately archives coordinates and then initiates the next sweep of HMC. Each sweep is initiated using the current coordinates and velocities assigned from a Gaussian distribution at the reciprocal temperature $\beta$. In assigning velocities, care is taken to ensure that the system does not have a net translational momentum.  The coordinates and velocities are then handed to the molecular dynamics
program. Here the molecular dynamics part of the simulation was performed using NAMD \cite{namd}. 

We first considered a system with 64 water molecules at a particle density of 
$\rho = 33.33$~{nm}$^{-3}$.  The initial oxygen coordinates were obtained
from the coordinates of a system of hard-spheres at a reduced density ($\rho \sigma^3$) of $0.3$. We purposely chose a poor initial configuration to estimate how well the $g_{\rm OO}(r)$ 
converges to that obtained using a Langevin dynamics simulation of a well-equilibrated system.  $N_{md} =10, 20, 50, 75,{\rm and}\, 100$ were considered. To compare with the {\em ab initio\/} simulations, we also considered a 32 particle system and 
$N_{md} = 50$. The initial configuration of the 32 particle system was obtained from an equilibrated configuration of SPC/E water molecules.

\subsection{{\em Ab initio} HMC simulations}
The molecular dynamics part of the simulations were performed using the publicly available {\sc cp2k}  code \cite{cp2knew}.
The HMC method was implemented as a script as discussed above. 

The {\sc cp2k} code uses  the Gaussian plane wave (GPW) method~\cite{cp2knew,GLippert97} based on the Kohn-Sham
 formulation of density functional theory together with a hybrid
 Gaussian and plane wave basis.  The norm-conserving pseudopotentials of
 Goedecker-Teter-Hutter~\cite{SGoedecker96,CHartwigsen98}
 (GTH) and a triple-$\zeta$ valance basis set augmented with two sets of
 {\it d}-type or {\it p}-type polarization function (TZV2P)  were used throughout; our choice follows
 several recent studies using the same basis \cite{kuo:jpc04,mcgrath:cpc05,joost:jcp05,mcgrath:vle06,kuhne:jctc09,mundy:jpcb09}.   
 A 280 Ry cutoff for the auxiliary plane 
 wave grid was employed, and the efficient and numerically stable orbital transformation energy minimizer
introduced in Ref.\ \cite{VWeber08b} is used to converge the SCF iterations to $10^{-6}$ a.u.\
of the Born-Oppenheimer surface. The nuclei are propagated by the velocity Verlet \cite{allen} algorithm. Standard masses
were used for hydrogen and oxygen. The simulation system comprises  $32$ water molecules at a number density of 
 33.33 nm$^{-3}$.  The initial configuration was obtained from an equilibrated 
 configuration of  SPC/E water molecules. 
  
The electronic structure was solved using the BLYP~\cite{BLYP:88,DFT_LYP:88}, revised PBE (revPBE)~\cite{YZhang98}, and
the BLYP-D \cite{Grimme:06} generalized gradient approximations to density functional theory. 
The BLYP-D functional includes an empirical correction (denoted by `-D') for dispersion interactions. 
Following a recent study \cite{mundy:jpcb09}, a  cutoff of 48{\AA} was used for 
the empirical dispersion contribution.

For all the functionals we report data using $N_{md} = 50$. (This choice is explained below. Test calculations
with revPBE and $N_{md} = 10, 20, 75$ show, as expected  \cite{mehlig:prb92}, the insensitivity of the 
results to choice of $N_{md}$.)

\subsection{Temperature effects}

In the present study we explore the classical statistical mechanics of liquid water, and an important element missing
from these studies is the effect of proton nuclear quantum effects on weakening intermolecular bonding. Earlier studies, for example Refs.\ \cite{rossky:jcp85,guissani:jcp98,schwegler:jcp04a,kusalik:jacs05},
of water using empirical interaction potentials find that increasing the temperature by about 50~K  mimics the effect of including proton nuclear quantum effects. A more recent study finds less dramatic quantum effects if molecular flexibility is 
considered and the parameters of the empirical model optimized with quantum effects \cite{manolopo:jcp09}. 

In this work, we regard temperature as a convenient parameter to change the {\em effective\/} strength 
--- $\beta\mathcal{U}$ is the pertinent quantity in sampling configurations --- of interaction obtained using a functional. 
In this study, we perform simulations at 300~K and 350~K, with the system simulated at the latter temperature 
mimicking the effect of  weaker intermolecular interactions.  Below, results using a 
particular functional and a given temperature are labelled by  `functional temperature' combination. 

\subsection{Statistical uncertainties}
Throughout this work,  statistical uncertainties were estimated following the block transformation procedure developed in
Allen and Tildesley \cite{allen}, following the earlier work of Friedberg and Cameron \cite{friedberg:1970}. In the case of the pair-correlation, each bin was treated as a separate channel for data and the error analysis was performed on the counts obtained in each channel. For obtaining uncertainties in $x_n$ and $p_n$, the appropriate number of instances per frame 
was used as the data stream and errors estimated. A similar approach was used for estimating uncertainties in 
$\langle \delta \mathcal{H}\rangle$ and $ \langle e^{-\beta\delta \mathcal{H}}\rangle$.

\section{Results and Discussion}\label{sc:results}

\subsection{HMC with SPC/Fw model and choice of $N_{md}$}
\begin{figure}
\begin{center}
\includegraphics{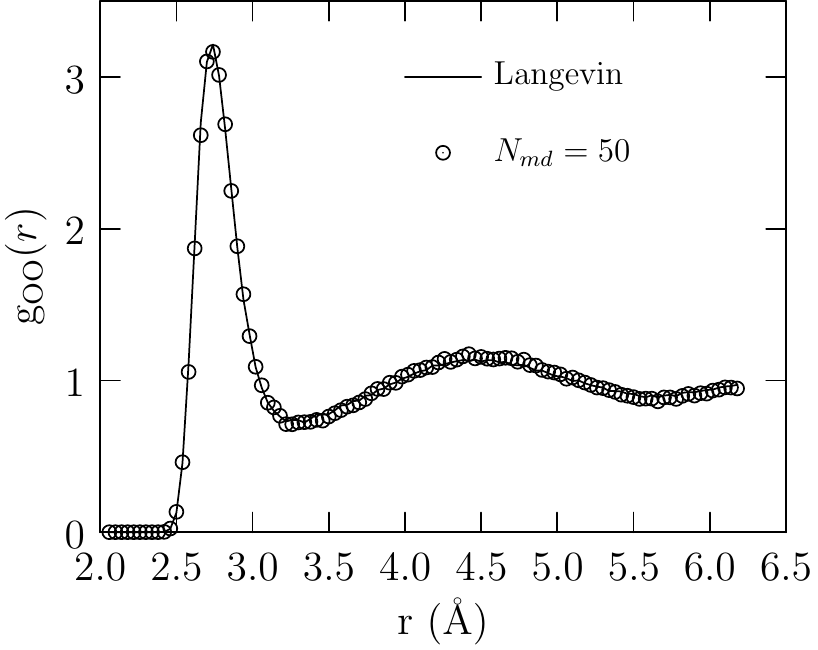}
\end{center}
\vspace{7cm}
\caption{Structure of SPC/Fw water at 300~K using the HMC and Langevin dynamics approaches. The results are
insensitive to the choice of $N_{md}$ and only results with $N_{md} = 50$ are shown. For the $N_{md} = 50$ 
simulation, $4\times 10^3$ sweeps were performed and the last $2\times 10^3$ sweeps were used for averaging. 
A fixed time step of $\delta t = 1$~fs was used. For the HMC simulation with $N=64$, the starting oxygen positions were based on those of a hard-sphere fluid at a reduced density of 0.3. Results using $N=32$ particles overlap those shown and are not included to preserve clarity. A bin-width of 0.04~{\AA} has been used for analyzing the data.}\label{fg:gr_spcfw}
 \end{figure}

Figure~\ref{fg:gr_spcfw} compares the structure of the SPC/Fw water obtained using the HMC 
and Langevin dynamics approaches. Results based on other choices of $N_{md}$ overlap those 
noted in Figure~\ref{fg:gr_spcfw} and are not shown for clarity.

A reasonable choice of $N_{md}$ can be inferred from the velocity autocorrelation time \cite{chandler:book} $\tau_{\rm vac} = \beta m \mathcal{D}$, where $m$ is the mass of the particle and $\mathcal{D}$ the diffusion coefficient. $\tau_{\rm vac}$ is the time by which velocities become uncorrelated and diffusive motion takes hold. Thus beyond $\tau_{\rm vac}$,
configurations are explored by a less efficient diffusive motion, a situation that we seek to avoid in the HMC. For liquid water, $\mathcal{D} \approx 0.25$~{\AA}$^2$/ps at 300~K \cite{mills:jpc73,krynicki:fd78} and $\tau_{\rm vac} \approx 200$~fs.

In the HMC procedure, by design, the velocities become uncorrelated every $N_{md}$ time steps, and for a 
fixed integration time-step ($\delta t$), the auto-correlation time $\tau_{\rm vac}$ is directly proportional to $N_{md}$. 
Thus $\delta t \cdot N_{md}$ very small compared to $\tau_{\rm vac}$ is akin to exploring configurational space diffusively
and is not to be preferred. On the other hand, a large $\delta t \cdot N_{md}$ can lead to larger integration errors ($\delta\mathcal{H}$) and lower acceptance rates. For these reasons, we conservatively choose $N_{md} = 50$ ($\approx \tau_{\rm vac} / 4$ for an integration time step of 1~fs) for all our simulations.  

\subsection{HMC with {\em ab initio\/} models}

In Table~\ref{tb:simdata} we collect several key metrics. Clearly, $\langle e^{-\beta\delta\mathcal{H}}\rangle \approx 1$, and  the value of the acceptance rate predicted by Eq.~\ref{eq:pa} is also close to the value actually found. Since $\delta t \approx 1$~fs, simulations with each density functional extended beyond about 170~ps. In the first 2000 sweeps (about 100~ps), $\delta t$ was varied to target an acceptance rate around 70\%. In the subsequent nearly 1400 sweeps (about 70~ps), $\delta t$ was held fixed. Of these, 500 sweeps (about 25~ps) were set aside for further equilibration and the remaining used in analysis. 

\begin{figure}
\begin{center}
\includegraphics{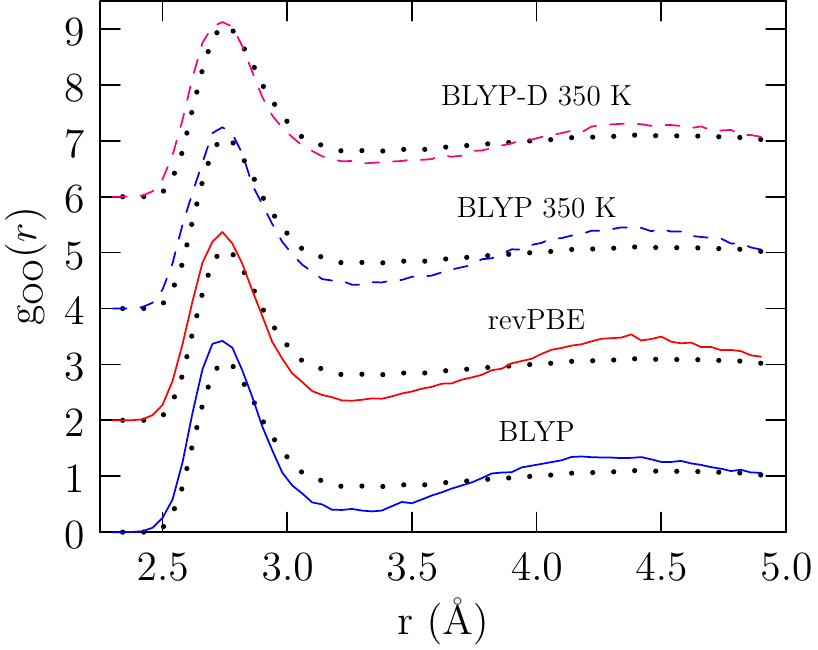}
\end{center}
\vspace{7cm}
\caption{Oxygen-oxygen radial distribution function at $300$~K (unless otherwise noted).
$g_{\rm OO}(r)$ for SPC/E (dotted line) is used as the standard for comparing different functionals. Data using SPC/Fw
is similar to that from SPC/E and is not shown. A bin-width of 0.04~{\AA} has been used for analyzing the data.}\label{fg:grall}
\end{figure}

\begin{figure*}
\begin{center}
\includegraphics{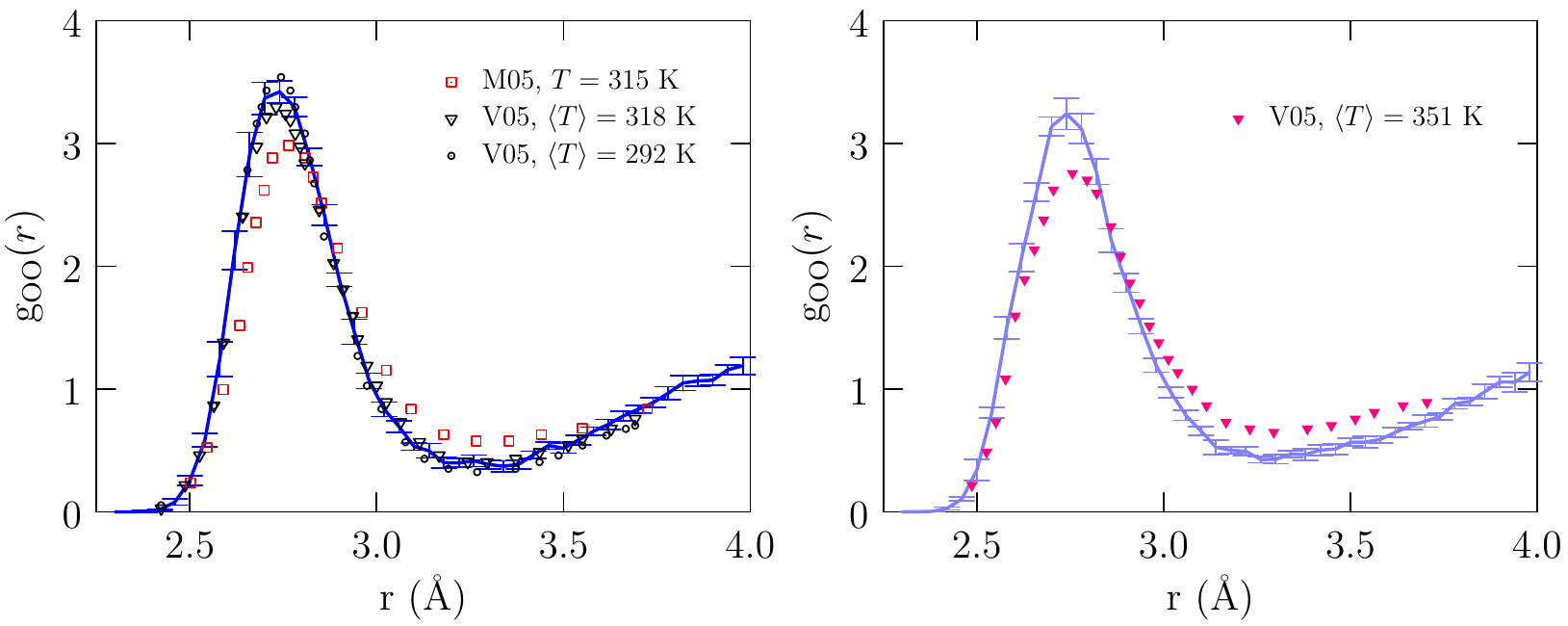}
\end{center}
\vspace{7cm}
\caption{Oxygen-oxygen radial distribution function using the BLYP density functional. {\em Right panel\/}: 
BLYP at 300~K (blue curve).  {\em Left panel\/}: BLYP at 350~K (light blue curve). The statistical uncertainties at 
the $2\sigma$ level are shown.  All calculations are based on the {\sc cp2k} code, the GTH-TZV2P basis set, and cutoff (280 Ry) for the charge density grid (except where noted). M05: Monte Carlo, with sampling of configurations 
using an empirical potential; a 1200 Ry cutoff for the charge density grid was used and the simulation comprises 64 water molecules  \cite{mcgrath:cpc05}. V05: $NVE$ Born-Oppenheimer molecular dynamics \cite{joost:jcp05} on a 32 water molecule system. Uncertainty of about $\pm$10~K was reported around the average temperature noted in the figure \cite{joost:jcp05}. A bin-width of 0.04~{\AA} has been used for analyzing the data.}\label{fg:grblyp}
\end{figure*}

In Figure~\ref{fg:grall} the radial density distribution of water oxygen is shown for the different functionals and temperatures
considered here, and in Figure~\ref{fg:grblyp} we compare the results of the current BLYP simulations with some of the earlier 
results based on the same ({\sc CP2K}) code and basis set (GTH-TZV2P). Comparing
BLYP 300~K and BLYP 350~K,  we find that at the higher temperature the first maximum is lowered by about 
0.2 (Table~\ref{tb:simdata}) and the first minimum is likewise elevated.  The change with temperature is in the right direction. 

 In comparing the present results to other simulation results,  some aspects need to be emphasized.  First, the response to temperature will sensitively depend on the water model (for example, see \cite{rossky:jcp85}).
Second, the response will sensitively depend on the simulation ensemble, especially when small system sizes are involved. With these caveats, observe that the location and magnitude of the first maximum for BLYP 300~K falls
between NVE ensemble results \cite{joost:jcp05} reported at  an average temperatures of  292~K and 318~K.  (The 
uncertainty in temperature was reported to be about $\pm$10~K around the average temperature \cite{joost:jcp05}.)  However, the agreement is less encouraging at 350~K (Figure~\ref{fg:grblyp}, right panel). The dependence of the thermal and 
the mechanical response of the liquid on the thermodynamic state point may underlie the observed difference. For example, 
the thermal expansion coefficient of water increases with temperature \cite{lrp:arxiv} and as does the compressibility
 beyond about 320~K.  (This likely also explains why in this study for revPBE~300~K a more structured pair-correlation function is  obtained in comparison to the earlier $NVE$ ensemble study \cite{asthagiri:pre03} of a 32 water molecule system  at a temperature of $314\pm 20$~K.)

Comparing the present results to those from an earlier Monte Carlo study, M05 in Figure~\ref{fg:grblyp} \cite{mcgrath:cpc05} (and also Ref.\cite{kuo:jpc04}), suggests that the peak of the correlation function obtained in that study is 
somewhat lower. (The location of the peaks are nearly the same.). Several factors
may underlie the observed differences.  First, the temperatures are different (Figure~\ref{fg:grblyp}).  Second, a larger system was used in the earlier study \cite{mcgrath:cpc05,kuo:jpc04}. The impact of system size will sensitively depend on the water model used. (We will return to this aspect below in discussing Figure~\ref{fg:pnscaled}.) 
For example, with the SPC/E water model, the pair-correlation approaches the bulk almost after the first hydration shell (Figure~\ref{fg:grall}) and both 32 and 64 particle simulation cells give essentially identical pair-correlation functions. But as 
Figure~\ref{fg:grall} suggests, the correlations appear more pronounced for water simulated with BLYP and  these correlations
can be expected to persist for longer length scales leading to more pronounced system size effects. 
Third, the different methodologies may be a factor. In the earlier Monte Carlo study \cite{mcgrath:cpc05,kuo:jpc04}, an empirical potential was used as an importance function to sample configurations\cite{gelb:jcp03}. (That empirical potential was parameterized against a Car-Parrinello simulation\cite{burnham:jcp04}.) But it is not clear if the empirical model was a good reference model\cite{wood:jcp99} for the target system studied. 

\subsection{Number distributions}

\begin{figure*}
\begin{center}
\includegraphics{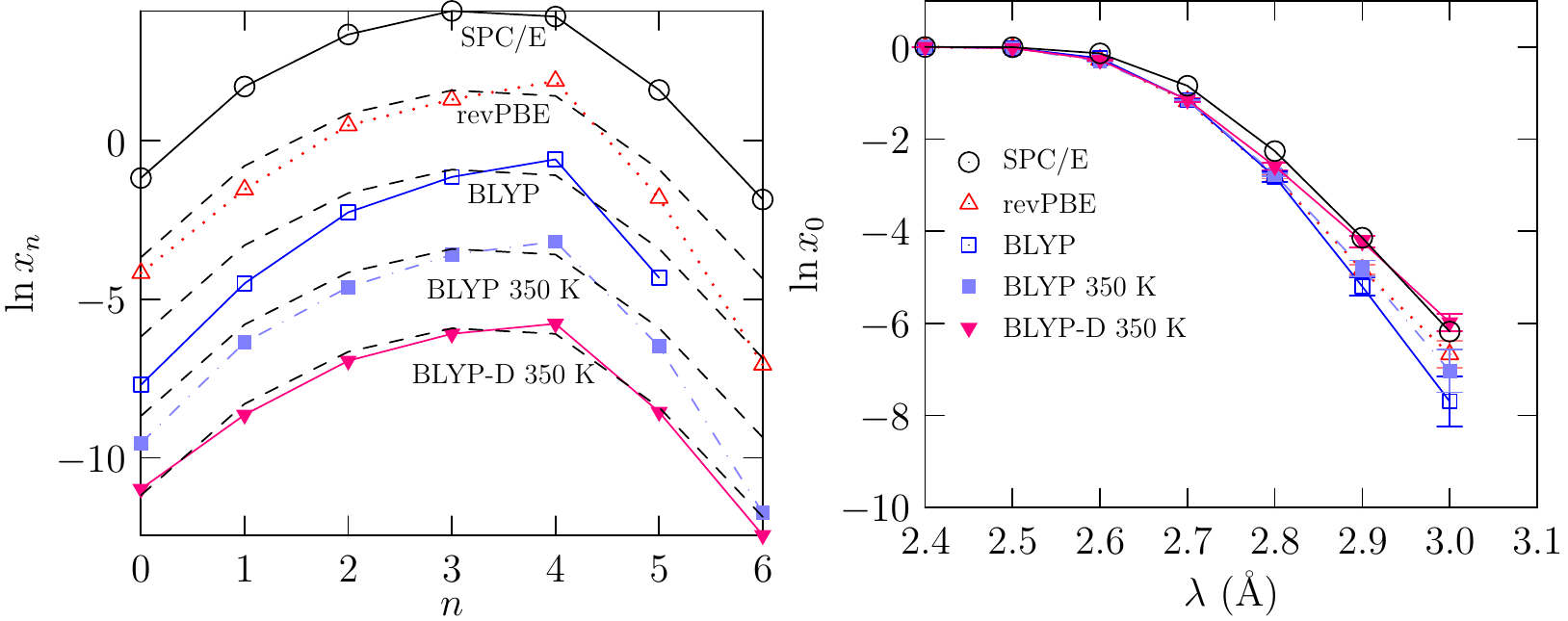}
\end{center}
\vspace{7cm}
\caption{{\em Left panel}: Distribution of coordination numbers around a distinguished water molecule. The coordination radius is 3.0~{\AA}.  The curves are translated vertically for clarity. The SPC/E $\{x_n\}$ distribution is overlain on the results from each of the  density functional to facilitate comparison.  {\em Right panel}:  Variation of the chemical contribution with coordination radius $\lambda$. Statistical uncertainty at the 1$\sigma$ level is shown.}\label{fg:xnx0}
\end{figure*}

Figure~\ref{fg:xnx0} (left panel) shows the distribution of coordination numbers observed around a distinguished water 
molecule. Fig.~\ref{fg:xnx0} (right panel) depicts the variation of the chemical contribution to the excess chemical potential 
for various coordination radii. 

For a chemically reasonable coordination radius, the coordination states below the most probable coordination 
state  reveal the importance of local interactions \cite{merchant:jcp09}.  As Figure~\ref{fg:xnx0} (left panel) suggests, relative to SPC/E,   the probability of the $n < 4$ states drops sharply  for BLYP 300~K. This 
suggests that BLYP 300~K leads to a somewhat tighter binding of the distinguished water molecule to the 
neighboring water molecules. If the local, cohesive interactions were weaker, we expect to
observe a greater proportion of the $n < 4$ states. Comparing  BLYP 300~K and BLYP 350~K shows 
that weakening the local, cohesive interactions does indeed elevate the proportion of  the $n < 4$ states. 
Comparing $\ln x_0$ for chemically meaningful coordination radii shows that the work of expelling water molecules
from within the coordination shell is more for BLYP 300~K than for any other case. 
Further, across all coordination states and coordination radii, the BLYP-D~350 K best follows the SPC/E results.

\begin{figure*}
\begin{center}
\includegraphics{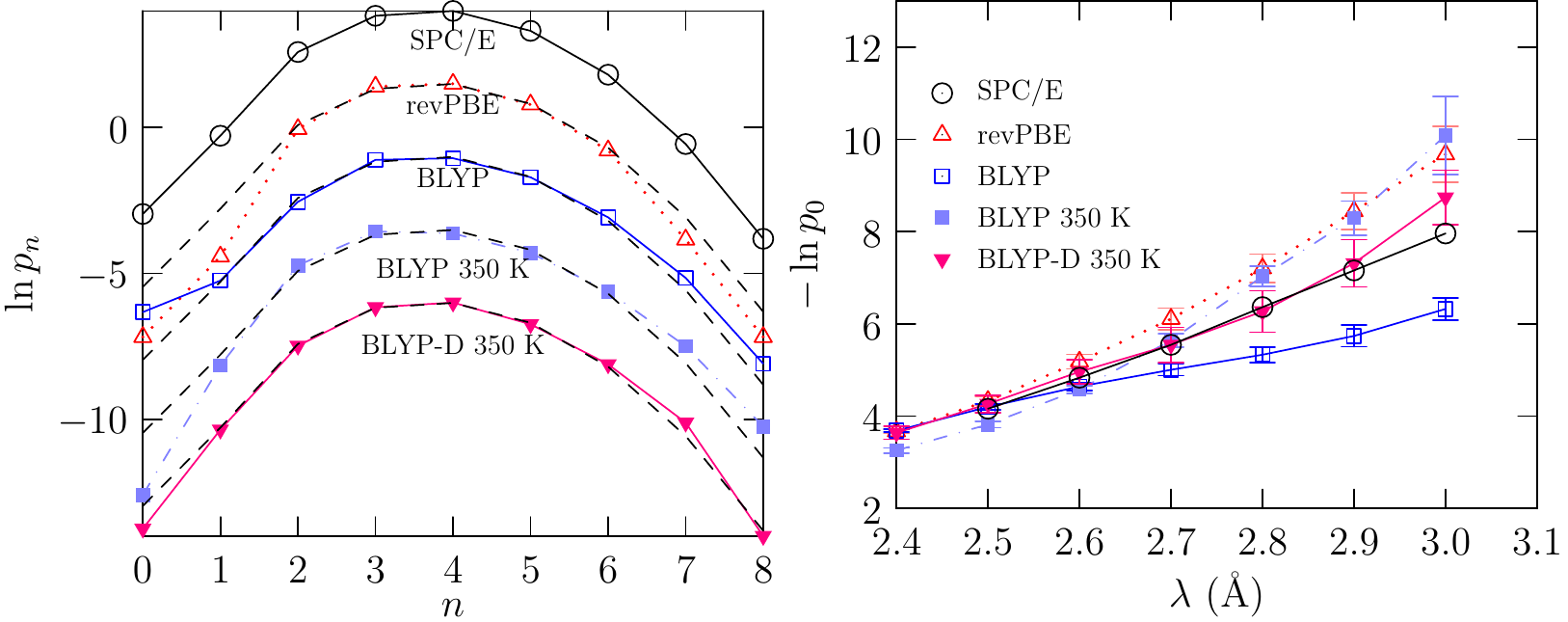}
\end{center}
\vspace{7cm}
\caption{{\em Left panel}: Distribution of water molecules in a cavity of radius 3~{\AA}.  The curves are translated vertically for clarity. The SPC/E $\{p_n\}$ distribution is overlain on the results from each of the  density functional to facilitate comparison.  {\em Right panel}:  Variation of the packing contribution with coordination radius $\lambda$. Statistical uncertainty at the 1$\sigma$ level is shown.}\label{fg:pnp0}
\end{figure*}

\begin{figure*}
\begin{center}
\includegraphics{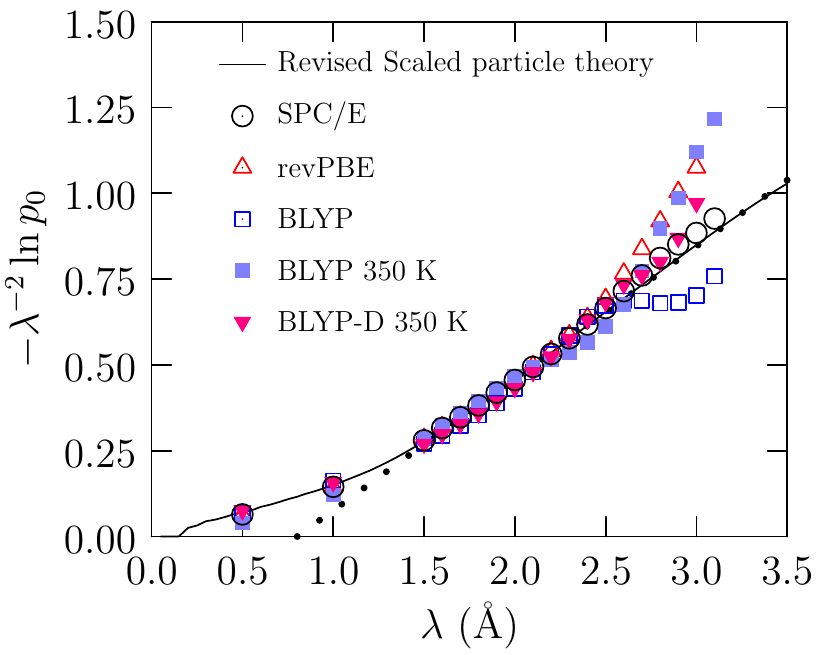}
\end{center}
\vspace{7cm}
\caption{The packing contribution scaled by the surface area (to within constants) versus the radius of the hard sphere. Compare also with Figure~2 in Ref.\cite{chandler:nature05}. The solid curve is based on the revised scaled particle
theory of Ashbaugh and Pratt \cite{ashbaugh:rmp}. A linear fit to the scaled particle results for $1.5\leq \lambda \leq 3.5$ is shown (dotted line).}\label{fg:pnscaled}
\end{figure*}

Figure~\ref{fg:pnp0} (left panel) gives the occupation statistics in an observation volume of radius 3~{\AA}. Comparing BLYP~300~K and BLYP~350~K shows that increasing the temperature makes it harder to open a cavity in liquid water. This is as expected, since the disorder in the medium increases with increasing temperature. (As an aside, this observation also explains why the solubility of hydrophobic  solutes decreases with increasing temperature \cite{lrp:jpcb98}.) Including additional attractive interactions mitigates the effect of increasing temperature, a trend more clearly seen in
the behavior of $\ln p_0$ (Figure~\ref{fg:pnp0}, right panel). 

\subsection{Scaling of the packing contribution}

To facilitate the discussion below, we first collect several observations about occupancy number distributions, the predicted 
scaling of the packing contribution based on theory, and number distributions and system size effects. 

In liquid water, for observation volumes of radii between about 1.5~{\AA} to 3.5~{\AA},  the occupation number distribution is found to closely approximate a Gaussian \cite{paliwal:jcp06,Hummer:1996p326,garde:prl96,lrp:jpcb98,chandler:nature05}. (The approximation is much better for the smaller radii.) Physically this means that at the size-scale of the observation volume, the presence of one molecule anywhere in the observation volume has only a modest (or little) effect in inducing the presence of another molecule in the volume.   But as we increase the size of the observation volume, more of the network structure of water  \cite{headgordon:pnas95,brovchenko:prl07,brovchenko:pccp07} comes into focus, and the occupation number is no longer Gaussian. In this instance, the presence of one water does induce the presence of another molecule in the volume.

Theory  \cite{garde:prl96,chandler:nature05} suggests that when the occupation number is approximately Gaussian, the
excess chemical potential of the empty observation volume (the cavity) scales with the volume of the cavity.  On the
other hand, the excess chemical potential of a large cavity depends on the surface area of the cavity. This scaling arises
because at the large length scale ($> 10$~{\AA}), the free energy is effectively the work it takes to create the interface. For the larger cavities, the preference for preserving hydrogen-bonding outweighs the need to accommodate the 
cavity \cite{chandler:nature05} in the liquid matrix. 

The occupation number distributions will be sensitive to system size effects. The number variation reflect positional 
fluctuations (of the molecules) and these will always be constrained in a constant volume simulation. In practice we find (Merchant and 
Asthagiri, in preparation) that system size effects manifest for observation spheres with a volume greater than 
about 3\% of the box volume. (For system sizes considered here, this amounts to an observation sphere of 
radius $\approx$ 2 {\AA}.) For cavities larger than this size, the excess chemical potential will be more positive 
than what would be obtained at constant pressure and/or large system sizes. The revised 
scaled particle theory \cite{ashbaugh:rmp} provides the packing contributions for cavities of various sizes in the
large system limit.

From Figure~\ref{fg:pnscaled}, we find that the packing contribution ($\beta\mu^{\rm ex}_{\rm HS} = -\ln p_0$) 
for BLYP~350~K and revPBE~300~K is above the scaled particle result for $\lambda > 2.4$~{\AA}. The
dependence of $-\beta \lambda^{-2} \mu^{\rm ex}_{\rm HS} $ on $\lambda$ is still linear, but the rate of increase is greater than that predicted theoretically. This indicates the onset of system size effects, and suggests that the pressure in the fluid is higher than normal. 
Thus in a constant temperature {\em and} pressure simulation of these systems, we would expect a lower density. A recent report \cite{mundy:jpcb09} using the BLYP functional at a temperature of 330~K and a pressure of 1~bar does indeed find a lower density for the fluid. The present results are consistent with that observation. (Although the temperature in that study was lower than the one here, we suspect that the above noted trend will hold.) Compared to both BLYP 350~K and revPBE~300~K, system size effects set in somewhat later in BLYP-D~350~K, just as they do for SPC/E.

The trend for BLYP is striking. For $\lambda$ between 2.5 to 3.0~{\AA}, the packing contribution scales with the 
surface area of the cavity, a scaling behavior that is not predicted to arise until after $\lambda \approx 10$~{\AA}.  (At that size
scale, this behavior reflects the need for water molecules to maximize their bonding by de-wetting the interface \cite{chandler:nature05}.) This unexpected behavior clearly reflects non-Gaussian occupation statistics (cf.\ Figure~\ref{fg:pnp0}, left panel), and suggests the presence of heterogeneities, perhaps strongly bonded pairs or other such clusters 
of water molecules, in the liquid.  (The more negative chemical contribution in BLYP 300~K water supports
this suggestion.) On the basis of temperature effects of water simulated with BLYP, an earlier report\cite{joost:jcp05}  found deviations of the diffusion coefficient from an Arrhenius behavior at temperatures around 300~K. Heterogeneities in the liquid can lead to such behavior and our results appear to corroborate the earlier study. A thorough analysis of the nature of heterogeneities/network structure of the liquid \cite{headgordon:pnas95,brovchenko:prl07,brovchenko:pccp07} will require a much larger system, preferably one where there are at least 3-4 hydration layers around a central water. Finally, 
for a somewhat larger cavity, $\lambda \geq 3.1$~{\AA},  $-\beta \lambda^{-2} \mu^{\rm ex}_{\rm HS} $ once again
begins to increase with $\lambda$, as must be expected, since the volume of the system is constant. 

\subsection{Balance of chemistry and packing}

\begin{figure*}
\begin{center}
\includegraphics{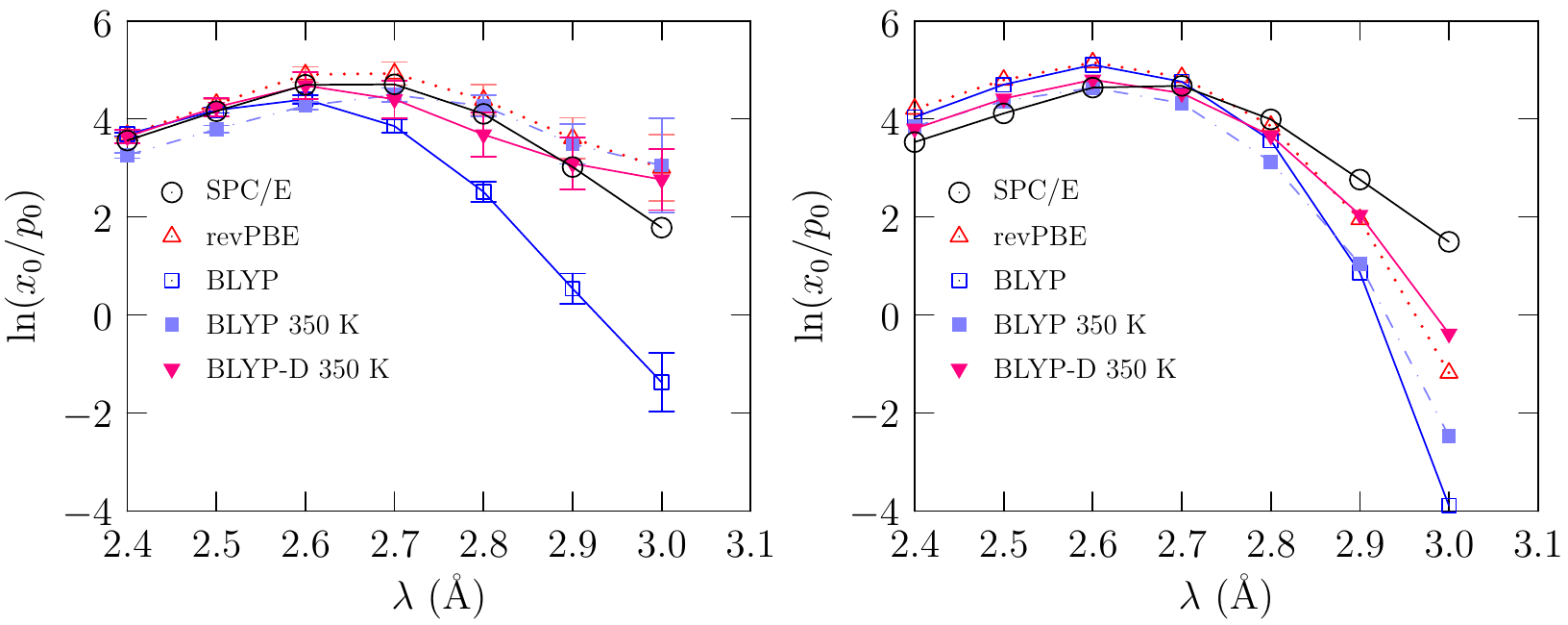}
\end{center}
\vspace{7cm}
\caption{{\em Left panel}: The sum of inner-shell chemistry and packing (steric) contributions to the excess free energy of hydration (in units of $k_{\rm B}T$). {\em Right panel}: The sum of inner-shell chemistry and packing {\em inferred} from a two moment information theory model, that is assuming Gaussian distribution of $\{x_n\}$ and $\{p_n\}$ \cite{Hummer:1996p326,garde:prl96,lrp:jpcb98,lrp:book}.}\label{fg:xopo}
\end{figure*}

Figure~\ref{fg:xopo} (left panel) shows the sum of the inner-shell chemistry and packing contributions obtained using the simulation data (Figs.~\ref{fg:xnx0} and~\ref{fg:pnp0}). Figure~\ref{fg:xopo} (right panel) are information theory\cite{asthagiri:pre03} estimates.
Consistent with the above discussion, we find that local chemistry outweighs packing for BLYP 300~K. As an {\em extreme}
approximation, if we assume that outer contribution is same for all the cases, then on the basis of the simulations
we expect that the excess chemical potential of water simulated with BLYP is about 4~${k_{\rm B}T}$s more negative than the
SPC/E value ($= -7.2$~kcal/mol, Merchant and Asthagiri, unpublished).  For all the other cases, relative to the 
SPC/E value, the excess chemical potential will be more positive by about a $k_{\rm B}T$.

Information theory predictions are only qualitatively consistent with the data; for example, BLYP is predicted to be more strongly bound than BLYP-D 350~K. However, the quantitative deviations from actual data can be as high as 5~$k_{\rm B}T$ 
(cf.\ BLYP~350~K, right and left panels of Fig.~\ref{fg:xopo}). But this is not surprising given that
$\{x_n\}$ (Fig.~\ref{fg:xnx0}, left panel) is non-Gaussian (see also\cite{paliwal:jcp06}, and the two-moment information theory model will only be approximate. 

\section{Conclusions}\label{sc:conclusions}

The free energy of expelling water molecules from the coordination sphere --- the chemical contribution to hydration --- of a distinguished water molecule  was calculated for liquid water simulated at 300~K with revPBE and BLYP functionals and at 350~K with BLYP and BLYP-D functionals. From this calculation we find that the distinguished water molecule is somewhat more tightly bound in water simulated with BLYP at 300~K than for the other cases. 

The hard-sphere packing contribution per unit surface area, $\beta\lambda^{-2}\mu^{\rm ex}_{\rm HS}$, of the hard-sphere was obtained for various radii ($\lambda$) of the hard-sphere and the results compared with those based on the revised scaled particle theory.  Except for  BLYP~300~K, $\beta\lambda^{-2}\mu^{\rm ex}_{\rm HS}$ scales 
linearly with $\lambda$, but shows distinct domains of linearity, a behavior consistent with the 
Gaussian occupancy statistics in an observation volume of the same size as the hard sphere.  

For revPBE~300~K and BLYP~350~K and $\lambda \geq 2.4$~{\AA}, $\beta\lambda^{-2}\mu^{\rm ex}_{\rm HS}$ increases faster than the scaled particle predictions and indicates the onset of system size effects earlier than it does for SPC/E. 
This shows that at density of 0.997~gm/cc (corresponding to 1~bar pressure in the SPC/E model), the pressure in these systems is higher than 1~bar. 

For BLYP~300~K,  $\beta\lambda^{-2}\mu^{\rm ex}_{\rm HS}$ is independent of $\lambda$ in the 2.4-3.0~{\AA} size-range. 
This behavior is not expected of liquid water at this size range and suggests the presence of heterogeneities in the medium.

\section*{Acknowledgments}
DA thanks the donors of the American Chemical Society Petroleum Research Fund for financial support. Financial support
for PD and DA from the National Science Foundation is gratefully acknowledged.  This research used resources of the National Energy Research Scientific Computing Center, which is supported by the Office of Science of the U.S. Department of Energy under Contract No. DE-AC02-05CH11231. 

\newpage

\newpage
\begin{table*}[h]
\caption{Statistics from the HMC simulations of water. The temperature $T = 1/(k_{\rm B}\beta)$. $N_{md} = 50$ is the number of molecular dynamics steps between consecutive Monte Carlo steps. $\tau$ is the total number of sweeps. In the first 2000 sweeps ($\approx 100$~ps) of {\em ab initio\/} simulations, $\delta t$ was periodically updated 
to target an acceptance rate of 0.7. Subsequently, $\delta t$ was held fixed at the value shown, and the first 500 sweeps of the total time shown were set aside for equilibration. For the SPC/Fw simulation, $\delta t=1$~fs was used always and the first 
2000 sweeps were set aside for equilibration. $\delta t$ is the time step for integrating the equations of motion. $\langle\beta\delta \mathcal{H}\rangle$ is the average discretization error between consecutive Monte Carlo steps.  The expected value of $ \langle e^{-\beta\delta \mathcal{H}}\rangle$ is 1. R is the ratio of the observed acceptance rate to that predicted by Eq.~\ref{eq:pa}. Statistical uncertainties at the $1\sigma$ level are noted, except for the $\rm g_{\rm OO}(r)$ where it is given at the
$2\sigma$ level. $r$  is the distance to the first maximum of  the $\rm g_{\rm OO}(r)$. Note that $r$ is defined to only within half the bin-width of 0.04~{\AA}. The simulation system comprises $32$ water molecules.}\label{tb:simdata}
\begin{tabular}{lcccccccc}
&   $\tau$ & $T$ (K) & $\langle \beta \delta \mathcal{H}\rangle$ & $ \langle e^{-\beta\delta \mathcal{H}}\rangle$ & $\delta t$ (fs) & R  & $r$ ({\AA}) & $\rm g_{\rm OO}(r)$  \\ \hline
SPC/Fw &  4000 & 300 & $0.20\pm 0.01$ & $1.00\pm0.01$ & 1.00 & 1.00 & 2.74 &  $3.21\pm0.06$ \\
BLYP &  1463 &  300 & $0.35\pm 0.06$ & $1.03 \pm 0.03$ & $1.16$ & 1.04 & 2.74 & $3.42\pm0.09$ \\
BLYP &  1459 &  350 & $0.31 \pm 0.03$ & $1.08 \pm 0.04$ & $1.12$ & 0.99 & 2.74 & $3.24\pm0.13$ \\ 
BLYP-D &  1400 & 350 & $0.26\pm 0.02$ & $1.01 \pm 0.03$ & $1.06$ & 0.97 & 2.74 & $3.12\pm 0.08$ \\ 
revPBE &  1391 &  300 & $0.26 \pm 0.03$ & $1.03 \pm 0.05$ & $1.02$ & 1.01 & 2.74 & $3.37\pm0.08$  \\ \hline
\end{tabular} 
\end{table*}

\newpage

\newpage

\newpage

\newpage

\newpage

\newpage

\newpage

\end{document}